\documentclass[%
reprint,
superscriptaddress,
showpacs,preprintnumbers,
nofootinbib,
 amsmath,amssymb,
 aps,
prb,
]{revtex4-1}

\usepackage{graphicx}% Include figure files
\usepackage{dcolumn}% Align table columns on decimal point
\usepackage{bm}% bold math

\begin{document}

%\preprint{}

\title{In-gap excitation effect on a superexchange in La$_2$CuO$_4$ by creating nonequilibrium photoexcited centers}



\author{Vladimir A. Gavrichkov}
%\thanks{}
\affiliation{L. V. Kirensky Institute of Physics, Siberian Branch of Russian Academy of Sciences, 660036 Krasnoyarsk, Russia}
%\affiliation{Siberian Federal University, 660041, Krasnoyarsk, Russia}
\author{Semen I. Polukeev}
%\thanks{}
\affiliation{L. V. Kirensky Institute of Physics, Siberian Branch of Russian Academy of Sciences, 660036 Krasnoyarsk, Russia}
%\affiliation{Siberian Federal University, 660041, Krasnoyarsk, Russia}

\author{Sergey G. Ovchinnikov}
%\thanks{}
\affiliation{L. V. Kirensky Institute of Physics, Siberian Branch of Russian Academy of Sciences, 660036 Krasnoyarsk, Russia}
\affiliation{National Research Nuclear University MEPHI (Moscow Engineering Physics Institute), Moscow 115409, Russia}


\date{\today}

\begin{abstract}
We propose a multielectron approach to calculate superexchange interaction in magnetic Mott-Hubbard insulator La$_2$CuO$_4$(further La214)
that allows to obtain the effect of optical pumping on the superexchange interaction.
We use the cell perturbation theory with exact diagonalization of the multiband $pd$ Hamiltonian
inside each CuO$_6$ unit cell and treating the intercell hopping as perturbation.
To incorporate effect of optical pumping we include in this work the excited single-hole local states
as well as all two-hole singlets and triplets. By projecting out the interband intercell electron hopping
we have obtained the effective Heisenberg-like Hamiltonian with the local spin at site $R_i$ being a superposition
of the ground and excited single-hole states. We found that antiferromagnetic contribution to the exchange energy in La214 will increase in accordance to {$\sim 4\cdot10^{-3} eV(\%)^{-1}$} at the resonance light occupation of the excited single hole in-gap state.

\end{abstract}
%\keywords{}

\maketitle


%\section{\label{sec:intr}Introduction\\}
\section{\label{sec:intr}Introduction\\}

Understanding the energy transfer between charge, orbital, and spin degrees of freedom is the important problem for many fields of solid state physics. Since the first experiments ~\cite{Beaurepaire_etal1996, Hohlfeld_etal1997} optical excitation of electronic spins and ultrafast magnetization dynamics have obtained  much attention. ~\cite{Stanciu_etal2007, Ostler_etal2012} A possibility to control the exchange interaction by light is important in many physics areas, from quantum computing~\cite{Duan_etal2003, Trotzky_etal2008, Chen_etal2011} to strongly correlated materials.~\cite{Wall_etal2009, Forst_etal2013, Li_etal2013} In many experiments the effect of optical pumping on the exchange interaction in the Mott-Hubbard insulators like manganites,~\cite{Wall_etal2009}  ferroborates,~\cite{Kalashnikova_etal2007, Kalashnikova_etal2008} TmFeO3, ErFeO3~\cite{Mikhaylovskiy_etal2014} etc. has been found. The origin of interatomic exchange interaction in all these oxides is related to the superecxhange mechanism via oxygen.~\cite{Anderson_1950} There are some simplified model calculations of the super exchange interaction under light irradiation in the three atomic model cation1-oxygen-cation2,~\cite{Moskvin_etal1979} that in complete theory should be extended to the crystal lattice.  The calculation of the superexchange interaction for the crystal lattice can be easily done for some simplified model like the Hubbard model.~\cite{Bulaevskii_etal1968, Chao_etal1977, Hirsch_1987} Within the LDA+DMFT approach the first-priciple calculations of the exchange interaction in correlated materials has been carried out in the work.~\cite{Katsnelson_etal2000} An idea of generalization of this approach to nonequilibrium optically excited magnetics has been also proposed in work.~\cite{Secchi_etal2013} without any practical conclusions. Nevertheless up to now the microscopic calculation of the superexchange interaction in La214 under light irradiation is absent.

%Depending on the frequency the light irradiation induces two types of optical transitions in the dielectric magnetic materials. The first type are the transitions with charge transfer through a dielectric gap and photocarrier generation.The second is intraionic optical transitions at a frequency lower than the radiation frequency corresponding to the fundamental absorption edge. Usually, it's the partially transparent and colored materials. There aren't any photocarriers.
%Under the optical irradiation the excited states of the cell are occupied.
%We use the hybrid ab initio  local density approximation (LDA) and  multielectron generalized tight binding (GTB) method that has been suggested earlier to calculated the electronic structure of

It is known that in the Hubbard model the superexchange $J$ results from the projecting out the interatomic hopping $t^{ab}$ accompanying the interband excitation from the low Hubbard band (LHB=a) to the upper Hubbard band (UHB=b). Due to the large insulator gap $U>>t^{ab}$ the interband excitation requires too much energy, and only virtual interband excitations from LHB to UHB and back are possible providing the exchange coupling $J \sim (t_{ab})^2/U$.~\cite{Bulaevskii_etal1968, Chao_etal1977, Hirsch_1987} The convenient mathematical tool for projecting out the irrelevant at large $U$ UHB is given by the projection operators.~\cite{Chao_etal1977}
%For realistic treatment of magnetic coupling in La$_2$CuO$_4$ within the LDA+GTB approach we have use similar projection operator method to calculates the exchange parameter $J$ in the ground state.~\cite{Gavrichkov_etal2008}

In this paper we calculate the exchange interaction in La214 under optical pumping within the hybrid LDA+GTB (generalized tight binding) approach.
Previously we have carried out similar calculation for La214 in the ground state.~\cite{Gavrichkov_etal2008} The LDA+GTB method allows to calculate the electronic structure of strongly correlated oxides like cuprates,~\cite{Korshunov_etal2005} manganites,~\cite{Gavrichkov_etal2010} boroxide~\cite{Ovchinnikov_2003, Ovchinnikov_Zabluda2004} and cobaltates.~\cite{Orlov_etal2013} We use the cell perturbation theory with exact diagonalization of the multiband $pd-$ Hamiltonian inside each CuO$_6$ unit cell with $ab~initio$ calculated parameters and treating the intercell hopping as perturbation. We restrict ourselves here by the antiferromagnetic undoped cuprate La214, nevertheless all ideas and methods used may be applied to any Mott-Hubbard insulator.
To incorporate effect of optical pumping we include in this work the excited single-hole local states as well as all excited two-hole singlets and triplets. It requires a generalization of the projection operators used here in comparison to the papers.~\cite{Chao_etal1977, Gavrichkov_etal2008}
Finally we have obtained the modification of exchange interaction induced by the light irradiation.
%We restrict ourselves by stationary irradiation, the time dependence of the demagnetization is outside the scope of this paper.
%\begin{widetext}

%\end{widetext}

\section{\label{sec:II}Effective superexchange hamiltonian\\}
\begin{figure}
\includegraphics{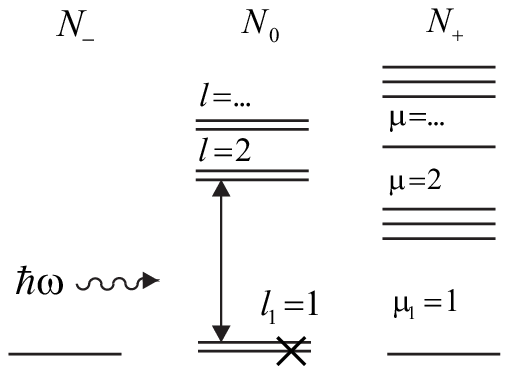}
\caption{\label{fig:1}}
\end{figure}

In the GTB approach one can assume that the quasiparticles are unit cell excitations which can be represented graphically as single-particle excitations (transitions) between different sectors $N_h=...(N_{0}-1), N_{0}, (N_{0}+1),...$ of the configuration space of the unit cell ($N_{0}$ is hole number per cell in the undoped material, see Fig.\ref{fig:1}).~\cite{Ovchinnikov_etal2012} Each of these transitions forms a $r$-th quasiparticle band, where the vector band index $r=\{i,i'\}$ in configurational space ~\cite{Zaitsev_1975} numerates the initial $i$ and final $i'$ many-electron states in the transition. The transitions, with the number of electrons increasing or decreasing, form the conduction or valence bands, respectively. For undoped La214 due to electroneutrality the proper subspace is $d9p6+d10p5$ with one hole per CuO$_6$ cluster, it has one hole, $N_0=1$. The hole addition requires $N_{+} = 2$ states $d9p5+d10p4+d8p6$. The hole removal results in $N_- =0$ states that for cuprates is given by a hole vacuum $d10p6$.
In the LDA+GTB method the Hamiltonian parameters are calculated ab initio\cite{Korshunov_etal2005} and the GTB cell approach
~\cite{Gavrichkov_etal1998,Ovchinnikov_etal2012} is used to take into account strong electron correlations explicity. A crystal lattice is divided into unit cells, so that the Hamiltonian is represented by the sum $H_0+H_1$, where the component $H_0$ is the sum of intracell terms and component $H_1$ takes into account the intercell hoppings and interactions. The component $H_0$ is exactly diagonalized. The exact multielectron cell states $|i\rangle$ ($|i'\rangle$) and energies $\xi_{i}$ are determined. Then these states are used to construct the Hubbard operators of the unit cell $\vec{R}_{f}: X^{i,i'}_f = |i\rangle\langle i'|$, where the meaning of the indexes $i$ and $i'$ is clear from Fig.\ref{fig:1}.

\begin{widetext}
\begin{equation}
H_0=\sum_{f}
\left\{%
\varepsilon_0X^{00}_f+\sum_{l\sigma}\left(\epsilon_l-\mu\right)X_f^{l\sigma,l\sigma}
+\sum_{\nu}^{N_\nu}(E_{\nu}-2\mu)X_f^{\nu,\nu} \right\}
\label{eq:1}
\end{equation}
\end{widetext}

is the sum of intracell terms and component $H_1$ takes into account the intercell hoppings and interactions. Here
\begin{eqnarray}
{H_1} &&= \sum\limits_{fg} {} \sum\limits_{\lambda \lambda '\sigma } {} t_{fg}^{\lambda \lambda '}c_{f\lambda \sigma }^ + {c_{g\lambda '\sigma }} + h.c. \nonumber \\
&&= \sum\limits_{fg} {} \sum\limits_{rr'} {} t_{fg}^{rr'}\mathop {X_f^r}\limits^ +  X_g^{r'}
\label{eq:2},
\end{eqnarray}
where $t_{fg}^{\lambda \lambda'}$ is the matrix of hopping integrals, and
\begin{eqnarray}
t_{fg}^{rr'}& = &
 \sum\limits_{\lambda \lambda '} \sum\limits_\sigma  t_{fg}^{\lambda \lambda '}\nonumber\\
 &\times &\left[ \gamma _{\lambda \sigma }^*\left( r \right)\gamma _{\lambda '\sigma }\left( r' \right) + \gamma _{\lambda '\sigma }^*\left( r \right)\gamma _{\lambda \sigma }\left( r' \right) \right],
 \label{eq:3}
\end{eqnarray}

where matrix elements:
\begin{eqnarray}
 {\gamma _{\lambda \sigma }}\left( {{r}} \right)& = &\left\langle ({{N_{+},{{M'}_{S'}}})_\mu} \right|c_{f\lambda \sigma } \left|({{N_{0},{M_S}})_l}\right\rangle\times \nonumber  \\
 & \times &\delta \left( {S',S \pm |\sigma| } \right)\delta \left( {M', M + \sigma } \right),
\label{eq:4}
\end{eqnarray}

Consideration is restricted by the case with one hole per cell $N_{0}=1$ in the undoped materials and an arbitrary number $N_\lambda$ of the occupied $\lambda$ orbitals, i.e. number of electrons $N_e=2N_\lambda-1$. This is relevant for the high-$T_c$ cuprates.
In this case of one hole per cell, the ${\left| {({N_{0}},{M_S})_i} \right\rangle} $ cell states are a superposition of different hole configurations of the same orbital (l) symmetry:
\begin{equation}
\left| ({N_{0},{M_{S}}})_{l} \right\rangle  = \sum\limits_\lambda ^{} {{\beta _l }\left( {{h_\lambda }} \right)\left| {{h_\lambda },{M_{S}}} \right\rangle }
\label{eq:5}
\end{equation}

Thus, there are one-hole spin doublet states, $C_{{2N_\lambda}}^{1} = {2N_\lambda}$, where  $C^k_n$ is the number of combinations. Besides, there are  $N_\mu=N_S+3N_T=C^2_{2N_\lambda}$ of the spin singlets $N_S=C_{N_\lambda}^{2}+N_\lambda$ and triplets $N_T=C_{N_\lambda}^{2}$:
\begin{equation}
\left| ({N_+,{M'_{S'}}})_{\mu} \right\rangle  = \sum\limits_{\lambda\lambda'} {{B_\mu}\left( {{h_\lambda},{h_{\lambda'}}} \right)\left| {{h_\lambda},{h_{\lambda'}},{M'_{S'}}} \right\rangle }
\label{eq:6}
\end{equation}
in the two-hole sector (Fig.\ref{fig:1}) in the ${N_\lambda}$ -orbital approach. Using the intracell Hamiltonian $H_0$ in the cell function  representation
the configuration weights $\beta_\mu(h_\lambda)$ and $B_\tau(h_\lambda,h_{\lambda'})$ can
 be obtained by the exact diagonalization procedure
 for the matrices $(\hat{H_0})_{\lambda\lambda'}$ and $(\hat{H_0})^{{\lambda\lambda'}}_{\lambda''\lambda'''}$
 in the $E_i({N_{h},{M_S}})$-eigenvalue problem  in different sectors $N_h$.~\cite{Ovchinnikov_etal2012}
 The sum (\ref{eq:2}) over all the $r$-th excited states with $l\neq l_1$ in the
 sector $N_0$  can not be omitted because of the light pumping.
 These excited states must be considered along with the $\mu$ - excited states in the  nearest $N_+$ sector.
%The component $H_0$ is exactly diagonalized. The exact multielectron cell states $|\mu\rangle$ ($|\nu\rangle$) and energies $\xi_{\mu}$ are determined. Then these states are used to construct the Hubbard operators of the $\vec{R}_{f}$ - unit cell  $X^{i,i'}_f $, where the meaning of the indexes $i$ and $i'$ is clear from Fig.\ref{fig:1}.

The superexchange interaction appears at the second
order of the cell perturbation theory with respect to hoppings.\cite{Jefferson_etal1992}
That corresponds to virtual excitations from the
occupied singlet and triplet bands through the insulating
gap to the conduction band and back.
These perturbations are described by the off-diagonal
elements $t_{fg}^{rr'}$ with $r=\{0l\}$ and $r'=\{l\mu\}$ in expression (\ref{eq:2}). In the Hubbard
model, there is only one such element $t^{01}$, which
describes the hoppings between the lower and upper
Hubbard bands.
In order to extract them, we generalize the
projection operator method proposed by Chao et al ~\cite{Chao_etal1977}
to the multiorbital GTB approach. Since the diagonal Hubbard
operators are projection operators, the $X$-representation
allows us to construct the set of projection operators. The total number of diagonal operators $X_f^{ii'}$
is equal to $N_\mu+N_l+1$ and the  sequence indexes $l$ and $\mu$ ($1 \le l  \le N_l$, $1 \le \mu  \le N_\mu$) runs over all electron states in the configuration spaces in Fig.\ref{fig:1}.
%The additions are to extend the approach~\cite{Chao_etal1977} to arbitrary configuration space of multielectron problem because all its states is under optical pumping. Because projective operators in work~\cite{Chao_etal1977}  are not additive by the number of excited states in $l-$ sector and  we can not write the completeness condition
%$\sum\limits_l  {p_l } = 1$,
%summing over all $l$ - ground and excited states. Nevertheless
Using a set of generalized operators

\begin{equation}
{p_0} = \left( {X_i^{00} + \sum\limits_l {X_i^{ll}} } \right)\left( {X_j^{00} + \sum\limits_l {X_j^{ll}} } \right),
\label{eq:7}
\end{equation}
and
\begin{equation}
{p_\mu } = X_i^{\mu \mu } + X_j^{\mu \mu } - X_i^{\mu \mu }\sum\limits_\nu  {X_j^{\nu \nu }}
\label{eq:8}
\end{equation}
with $\mu(\nu) =1,2,...N_\mu$ we can identify the contribution to the superexchange from the interband transitions. As will be seen below, a generalized approach with the operators (\ref{eq:7}) and (\ref{eq:8})  differs from the work~\cite{Chao_etal1977} just in details.
It can be checked that each of operator ${p_0 }$ and  ${p_\mu }$ is a projection
operator $p_0^2=p_0$ and $(p_\mu ^2 = {p_\mu })$. These operators also form a complete and orthogonal system, $p_0+\sum\limits_{\mu  = 1}^{N_\mu} {{p_\mu }}  = 1$, ${p_0 }{p_\mu } =0$ and ${p_\mu }{p_\nu } = {\delta _{\mu \nu }}{p_\mu }$. We highlight the diagonal and off-diagonal matrix elements
in expression:
\begin{eqnarray}
&H& =(H_0+H_1^{in})+H_1^{out},
\label{eq:9}
\end{eqnarray}
According to the work we introduce a Hamiltonian of the exchange-coupled $(ij)$-th pair: $h=(h_0+h_1^{in})+h_1^{out}=H_{ij}$,
where $H=\sum\limits_{ij}H_{ij}$ and
\begin{eqnarray}
h_0+h_1^{in} ={p_0}h{p_0}+\sum\limits_{\mu \nu } {{p_\mu}h{p_\nu}}
\label{eq:10}
\end{eqnarray}
and
\begin{eqnarray}
h_1^{out} ={p_0}h\left(\sum\limits_{\mu }{p_\mu}\right)+\left(\sum\limits_{\mu} {p_\mu}\right)h{p_0}
\label{eq:11}
\end{eqnarray}
are intra- and inter- band contributions in $H_1$ respectively.
%Only the latter of which  are involved in the superexchange interaction.
We perform the standard unitary transformation to project out the interband hopping and to derive superexchange interaction
\begin{equation}
{\tilde{h}} = {e^{G}}h{e^{-G}},
\label{eq:12}
\end{equation}
 where the matrix $\hat G$ satisfies to the equation
\begin{eqnarray}
{p_0}{h}\left(\sum\limits_{\mu}{p_\mu }\right)&+&\left(\sum\limits_{\mu}{p_\mu }\right){h}{p_0}+\nonumber\\
&+&{\left[ {{G,\left({p_0}h{p_0}+\sum\limits_{\mu \nu } {{p_\mu}h{p_\nu}}\right)}} \right] } = 0,
\label{eq:13}
\end{eqnarray}
and transformed Hamiltonian are given by
%\begin{widetext}
\begin{eqnarray}
\tilde{h}&\approx&\left({p_0}h{p_0}+\sum\limits_{\mu \nu } {{p_\mu}h{p_\nu}}\right)+ \nonumber\\
&+&\frac{1}{2}\left[ {{G,\left({p_0}{h}\sum\limits_{\mu}{p_\mu }+\sum\limits_{\mu}{p_\mu }{h}{p_0}\right)}} \right]
 \label{eq:14}
\end{eqnarray}
%\end{widetext}

where the contributions from inter-band transitions can be calculated as:
\begin{equation}
{p_0}{h}\left(\sum\limits_{\mu}{p_\mu }\right) =\sum\limits_{ll'\mu} {t_{ij}^{l0,l'\mu }X_i^{l0}X_j^{l'\mu }}
\label{eq:15}
\end{equation}
\begin{equation}
\left(\sum\limits_\mu  {{p_\mu }}\right) {h}{p_0} =  {\sum\limits_{ll'\mu } {t_{ij}^{\mu l',0l}X_i^{\mu l'}X_j^{0l}} }
\nonumber
\end{equation}
 Note, due to the absence of additivity over $l$-number of the excited state in the projective operator $p_0$,
the solution of Eq.(\ref{eq:13})  has the form

\begin{equation}
G = \sum\limits_\mu  { {\sum\limits_{ll'} {\frac{{t_{ij}^{l0,l'\mu }}}{{{\Delta _{ll'\mu }}}}} \left( {X_i^{\mu l'}X_j^{0l} - X_i^{l0}X_j^{l'\mu }} \right)} }
\label{eq:16}
\end{equation}
where $\Delta_{ll'\mu}=\varepsilon _0 + \varepsilon _\mu - \left( \varepsilon _l + \varepsilon _{l'}\right)$,
and the commutator in (\ref{eq:14}) can be represented as
\begin{widetext}
\begin{eqnarray}
    \delta h=\frac{1}{2}&&\sum\limits_{\mu \nu } {\left\{ {\left[ {{G_\nu },\left( {{p_0}{h}{p_\mu } +{p_\mu }{h}{p_0}} \right)} \right]} \right\}}=
 \nonumber \\
 && = \frac{1}{2}\sum\limits_{\mu \nu } {\left\{ {\left[ {\sum\limits_{ll'} {\frac{{t_{ij}^{l0,l'\nu }}}{{{\Delta _{ll'\nu }}}}\left( {X_i^{\mu l'}X_j^{0l} - X_i^{l0}X_j^{l'\mu }} \right)} ,\sum\limits_{kk'} {t_{ij}^{k0,k'\mu }\left( {X_i^{0k}X_j^{\mu k'} + h.c.} \right)} } \right]} \right\}}.
    \label{eq:17}
\end{eqnarray}
\end{widetext}
The right part of the exprexion (\ref{eq:14}) for
effective Hamiltonian $\tilde h$ can now be derived explicity. Calculating commutator in the above expression (\ref{eq:17}) hence we obtain the effective Hamiltonian for the exchange-coupled $(ij)$-th pair as

\begin{widetext}
\begin{eqnarray}
\delta \tilde{h} = \sum\limits_{ll'kk'} {\sum\limits_{\mu \nu } {\left( {\frac{{t_{ij}^{l0,l'\nu }t_{ij}^{k0,k'\mu }}}{{{\Delta _{ll'\nu }}}}} \right){\delta _{\mu \nu }}\left\{ {X_i^{l \uparrow ,k \uparrow }X_j^{l' \downarrow ,k' \downarrow } + X_i^{l \downarrow ,k \downarrow }X_j^{l' \uparrow ,k' \uparrow } - \left( {X_i^{l \uparrow ,k \downarrow }X_j^{l' \downarrow ,k' \uparrow } + X_i^{l \downarrow ,k \uparrow }X_j^{l' \uparrow ,k' \downarrow }} \right)} \right\}} }  &+& \nonumber \\
 +{\sum\limits_{ll'kk'}\sum\limits_{\mu \nu } {\left( {\frac{{t_{ij}^{l0,l'\nu }t_{ij}^{k0,k'\mu }}}{{{\Delta _{ll'\nu }}}}} \right) {{\delta _{kl}}{\delta _{k'l'}}\left( {X_i^{00}X_j^{\mu \nu } + X_i^{\mu \nu }X_j^{00}} \right) } } }=\delta {\tilde{h}_{s-ex}}+ \delta {\tilde{h}_\rho }&&,\nonumber \\
\label{eq:18}
\end{eqnarray}
\end{widetext}
and only a first contribution includes superexchange interaction $\delta {H_{s - ex}} = \sum\limits_{ij}\tilde{h}_{s-ex}$:

\begin{widetext}
\begin{eqnarray}
\delta {H_{s - ex}} = \sum\limits_{ij} {\sum\limits_{ll'kk'} {\sum\limits_\mu  {\frac{2{\left( {t_{ij}^{l0,l'\mu }t_{ij}^{k0,k'\mu }} \right)}}{{{\Delta _{ll'\mu }}}}\left\{ {\left( {{\delta _{{l_1}k}}Z_{il}^ +  + {\delta _{{l_1}l}}Z_{ik}^ +  + {\delta _{lk}}{{\hat S}_{il}}} \right)\left( {{\delta _{{l_1}k'}}Z_{jl'}^ +  + {\delta _{{l_1}l'}}Z_{jk'}^ +  + {\delta _{l'k'}}{{\hat S}_{jl'}}} \right)} \right.} } }&&  - \nonumber \\
\left. { - \frac{1}{4}\left( {{\delta _{{l_1}k}}y_{il}^ +  + {\delta _{{l_1}l}}y_{ik}^ -  + {\delta _{lk}}{n_{il}}} \right)\left( {{\delta _{{l_1}k'}}y_{jl'}^ +  + {\delta _{{l_1}l'}}y_{jk'}^ -  + {\delta _{l'k'}}{n_{jl'}}} \right)} \right\}&&,\nonumber\\
\label{eq:19}
\end{eqnarray}
\end{widetext}
where $S_{il}^ +  = X_i^{l\uparrow,l\downarrow }$, $2S_{il}^z = \sum\limits_\sigma  {\eta \left( \sigma  \right)X_i^{{l}\sigma ,{l}\sigma }}$, $Z_l^+= \hat S_{il_1}X_i^{l_1l}$ and $y_{il}^ +  = {\hat n_{i{l_1}}}X_i^{{l_1}l}$ are a spin--exciton and electron-exciton operators at the $i$-th cell. For simplicity, we assumed that $X_i^{l\sigma,l'\sigma}=X_i^{l\bar{\sigma},l'\bar{\sigma}}=X_i^{l,l'}$.
 Note that the contribution in Eq.(\ref{eq:19}) at $l=k$ and $l'=k'$
\begin{eqnarray}
 \delta H_s= \sum\limits_{ij} {\sum\limits_{ll'} {\sum\limits_\mu  {\frac{{{{2\left( {t_{ij}^{l0,l'\mu }} \right)}^2}}}{{{\Delta _{ll'\mu }}}}\left({S_{il}}{S_{jl'}}-\frac{1}{4}n_ln_{l'}\right)} } },  \nonumber \\
  \label{eq:20}
 \end{eqnarray}
 where $ {\hat S_{il}}{\hat S_{jl'}} = \frac{1}{2}\sum\limits_\sigma  {\left( {X_i^{l\sigma l\bar \sigma }X_j^{l'\bar \sigma l'\sigma } - X_i^{l\sigma l\sigma }X_j^{l'\bar \sigma l'\bar \sigma }} \right)} $,  is an analogue of the conventional superexchange with exchange constant $J_{ij}^{ll'} = 2\sum\limits_\mu  {{{{{\left( {t_{ij}^{l0,l'\mu }} \right)}^2}} \mathord{\left/
 {\vphantom {{{{\left( {t_{ij}^{l0,l'\mu }} \right)}^2}} {{\Delta _{ll'\mu }}}}} \right.
 \kern-\nulldelimiterspace} {{\Delta _{ll'\mu }}}}} $.

An exciton energy
 can not exceed the semiconductor gap ${E_g} = \left[ {{\varepsilon _{{\mu _0}}} + {\varepsilon _0} - 2{\varepsilon _{{l_1}}}} \right]$, because of the divergence of superexchange contributions $\delta H_{s-ex}\rightarrow\infty$ at $\delta_{ll_1}\rightarrow E_g$. At $\delta _{l{l_1}}>E_g$ the exciton cell state decays into an electron-hole pair state. Therefore photocarriers are generated under light pumping with a frequency $h\nu_q$ higher than the absorption edge, and the superexchange  on the photoexcited intracell states can be calculated in approach (\ref{eq:19})  only at the light pumping with the frequency in the transparency region of the material. It's partly colored magnetic nondoped materials.
%\begin{widetext}
\begin{figure*}
\includegraphics{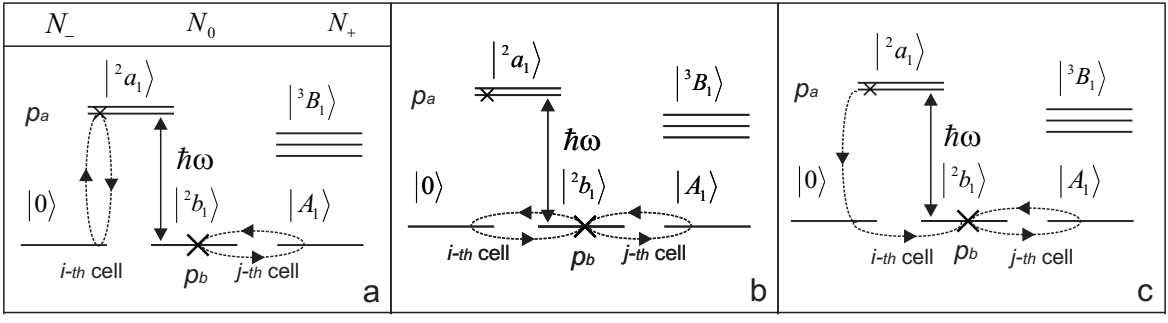}
\caption{\label{fig:2}Two circles (dashed line) are a sequence of intracell transitions at the light-induced superexchange $J_{ij}^{ab}$ ($a$) and $J_{ij}^{bb}$ ($b$) between $i$ and $j$ cells in Eq.(\ref{eq:25}), ($c$) illustrates the single circle (spin-exciton) contribution $\sim\left( {t_{ij}^{a0,bA }t_{ij}^{b0,bA }} \right)/{{{\Delta _{abA }}}} $ (see Eq.(\ref{eq:19})) which can not be reduced to the spin Hamiltonian  $\delta H_s$. }
\end{figure*}
%\end{widetext}

 Let's obtain the contribution (\ref{eq:20}) to the exchange energy of the system in the framework of mean-field approximation.
 \begin{widetext}
 \begin{eqnarray}
 \delta {H_{s - ex}} =  - \frac{1}{2}\sum\limits_{ij} {\sum\limits_{ll'} {J_{ij}^{ll'}} \left\langle {X_i^{l\sigma l\sigma }} \right\rangle \left\langle {X_j^{l'\bar \sigma l'\bar \sigma }} \right\rangle }  \approx  - \frac{{zN}}{2}\left[ {J_{\left\langle {ij} \right\rangle }^{{l_1}{l_1}}p_{{l_1}}^2 + 2\sum\limits_{l \ne {l_1}} {J_{\left\langle {ij} \right\rangle }^{l{l_1}}{p_l}{p_{{l_1}}} + } \sum\limits_{l,l' \ne {l_1}} {J_{\left\langle {ij} \right\rangle }^{ll'}{p_l}{p_{l'}}} } \right]
 \label{eq:21}
\end{eqnarray}
\end{widetext}
where ${p_{{l_1}}} = 1-{\left(\sum\limits_{l \ne {l_1}}{p_l}  {} \right)} $ and ${p_l} = \left\langle {X_i^{l \uparrow l \uparrow }} \right\rangle  = \left\langle {X_i^{l \downarrow l \downarrow }} \right\rangle $ is a probability to detect a cell in the excited state $\left| {{{\left( {{N_0},{M_S}} \right)}_l}} \right\rangle $.
%We also used the relation $(zN){J^{ll'}}\sim \sum\limits_{<ij>} {J_{ij}^{ll'}} $ and following expression:
%\begin{equation}
%p_l=
%\end{equation}
%In accordance with the Fermi golden rule probability $<Z_l^{\pm}>$ at $t$ longer than the characteristic time $t>{h}/{E_g}(\sim10^{-16}c\div10^{-15}c) $ of the optical transition:
%\begin{eqnarray}
%<Z_l^ +>  \sim t{n_q}<S_{l_1}>\delta \left( h\nu_q - \delta_{ ll_1}  \right)
% \label{eq:22}
%\end{eqnarray}
%and
%\begin{eqnarray}
%<Z_l^ ->  \sim t\left( {{n_q} + 1} \right)<S_{l_1}>\delta \left( {h{\nu _q} - {\delta _{l{l_1}}}} \right)
%\label{eq:23}
%\end{eqnarray}
%where ${\delta _{l{l_1}}} = {\varepsilon _l} - {\varepsilon _{{l_1}}}$ and ${n_q}={\left\langle {a_q^ + {a_q}} \right\rangle }$ are  the exciton energy and photon number in $q$-mode of the irradiation with Hamiltonian ${H_{ir}} = \sum\limits_q {h{\nu _q}\left( {a_q^ + {a_q} + \tfrac{1}{2}} \right)} $.
Thus  the light pumping effects  in  superexchange are frequency selective and linear on the amplitude pumping. In compound La214 the ground cell state is formed by a single hole $b_1$ orbital, the $a_1$ orbital may be excited by the external pumping (Fig.\ref{fig:2}).The standard mechanism of the superexchange in the ground state is shown in Fig.\ref{fig:2}b, while the superexchange via optically excited term is shown in Fig.\ref{fig:2}a. the formation of spin-exciton interaction that is beyond the Heisenberg model is shown in Fig.\ref{fig:2}c.

\section{\label{sec:III} Results for  copper oxide $La214$}

 We test the approach on the  high-$T_c$ parent materal La214. At the LDA parameters of Hamiltonian taken from ~\cite{Korshunov_etal2005} $ J_{bb}\approx$0.15 $eV$, $\delta_{ll_1}=\delta_{ab}$=1.78 $eV$, $E_g=$2.00 $eV$, and the $r=\{^2b_1,A_1\}$ - band index ~\cite{Feiner_etal1996, Gavrichkov_etal2001} corresponds to  $\{l_1,\mu=1\}$ first removal electron state.

Using the exact diagonalization procedure with LDA parameters, one obtains the weights $\alpha_l$, $\beta_l$ and $A_\mu$, $B_\mu$ at the doublet and singlet, triplet states:
\begin{widetext}
\begin{equation}
\left| ({N_{0},{M_{S}}})_{l=1} \right\rangle  = {\left| {{}^2{b_1}} \right\rangle } = \sum\limits_{\lambda  = {d_z},{p_z},a} {{\beta _{{l=1}}}\left( {{h_\lambda }} \right)\left| {{h_\lambda },{\sigma _{\tfrac{1}{2}}}} \right\rangle } ; \left| ({N_{0},{M_{S}}})_{l=2} \right\rangle  ={\left| {{}^2{a_1}} \right\rangle } = \sum\limits_{\lambda  = {d_z},{p_z},a} {{\alpha _{l=2}}\left( {{h_a}} \right)\left| {{h_a},{\sigma _{\tfrac{1}{2}}}} \right\rangle },
 \label{eq:23}
\end{equation}
\end{widetext}

\begin{widetext}
\begin{eqnarray}
\left| ({N_+,{M'_{S'}}})_{\mu=1} \right\rangle  &=& {\left| {{A_1}} \right\rangle } = \sum\limits_{\lambda ,\lambda ' = b,{d_x},a,{p_{z,}}{d_z}} {A_{\mu=1} }\left( {{h_\lambda },{h_{\lambda '}}} \right)\left| {{h_\lambda },{h_{\lambda '}},{0}} \right\rangle , \nonumber \\
\left| ({N_+,{M'_{S'}}})_{\mu=2} \right\rangle  &=& {\left| {{}^3{B_1}} \right\rangle } = \sum\limits_{\lambda  = b,{d_x}} {\sum\limits_{\lambda ' = a,{p_z},{d_z}} {{B_{\mu=2} }\left( {{h_\lambda },{h_{\lambda '}}} \right)\left| {{h_\lambda },{h_{\lambda '}},{M_1}} \right\rangle } },
\label{eq:24}
\end{eqnarray}
\end{widetext}
where $h_b$ and $h_{d_x}$ are the holes in the $b$-symmetrized cell states of oxygen and $d_{x^2-y^2}$ cooper states of the CuO$_2$ layer, respectively.

Because of $ \delta_{ab}<E_g$, just two contributions from the doublets ${\left| {{}^2{a_1}} \right\rangle }$ and ${\left| {{}^2{b_1}} \right\rangle } $ are available in the sum (\ref{eq:18}) over $l$.
Due to the symmetry CuO$_2$ layer $\gamma_\lambda(\{^2a_1,A_{1}\})=0$ at any $\lambda$ and therefore  $t_{ij}^{{b}0,aA }=t_{ij}^{{a}0,aA }=0$. Thus we evaluate the contribution (\ref{eq:23})
like the next:
\begin{widetext}
\begin{eqnarray}
\left\langle {\delta {H_{s - ex}}} \right\rangle  =  - \frac{{zN}}{2}\sum\limits_\mu  {\left[ {\frac{{{{\left( {t_{}^{b0,b\mu }} \right)}^2}}}{{{\Delta _{b\mu }}}}p_b^2 + 2\left( {\frac{{{{\left( {t_{}^{b0,a\mu }} \right)}^2}}}{{{\Delta _{ba\mu }}}} + \frac{{{{\left( {t_{}^{a0,b\mu }} \right)}^2}}}{{{\Delta _{ba\mu }}}}} \right){p_a}{p_b} + \frac{{{{\left( {t_{}^{a0,a\mu }} \right)}^2}}}{{{\Delta _{b\mu }}}}p_a^2} \right]} \sim &&\nonumber \\
\sim  - \frac{{zN}}{2}  {\left[ {0.15(eV)\cdot p_b^2 +2 \frac{{{{\left( {t_{}^{a0,bA_1}} \right)}^2}}}{{{\Delta _{baA_1}}}}{p_a}{p_b}} \right]}&&
\label{eq:25}
\end{eqnarray}
\end{widetext}

 \begin{figure}
\includegraphics{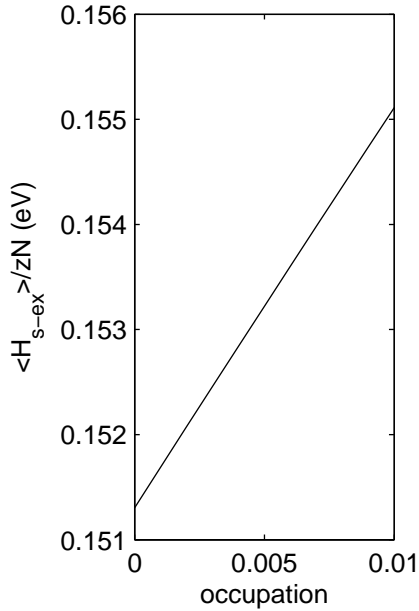}
\caption{\label{fig:3} A dependence of the antiferromagnetic contribution (\ref{eq:25}) on the occupation $p_a$ of excited state $|^2a_1>$}
\end{figure}

Without external irradiation $p_b=1$, $p_a=0$, and Eq.(\ref{eq:25}) results in the exchange interaction $J_{bb}$ (the first term in the right side of Eq.(\ref{eq:25})) in the ground state
obtained earlier in the work.~\cite{Chao_etal1977}
%we can further simplify the expression for the super exchange:
%\begin{widetext}
%\begin{eqnarray}
%J_1^{ij}= \left(w^+_a+w^-_a\right)\sum\limits_\mu  {t_{ij}^{{b}0,{b}\mu }}
% {\left( {\frac{{t_{ij}^{a0,{b}\mu }}}{{{\Delta _{{b}a\mu }}}} + \frac{{t_{ij}^{{b}0,a\mu }}}{{{\Delta _{{b}a\mu }}}}} \right)}\sim \left(w^+_a+w^-_a\right)\sum\limits_\mu
% {\left( {\frac{{{t_{ij}^{{b}0,{b}\mu }}t_{ij}^{a0,{b}\mu }}}{{{\Delta _{{b}a\mu }}}} } \right)}
% \label{eq:25}
%\end{eqnarray}
%\end{widetext}
%and
%\begin{widetext}
%\begin{eqnarray}
%{J_2^{ij}} = \sum\limits_\mu  {\left\{ {{{\left( {w_a^ + } \right)}^2}\frac{{t_{ij}^{{b}0,{b}\mu }t_{ij}^{a0,a\mu }}}{{{\Delta _{{b}\mu }}}} + \left( {w_a^ + w_a^ - } \right)\frac{{{{\left( {t_{ij}^{a0,b\mu } + t_{ij}^{b0,a\mu }} \right)}^2}}}{{{\Delta _{{b}a\mu }}}} + {{\left( {w_a^ - } \right)}^2}\frac{{t_{ij}^{{b}0,{l_0}\mu }t_{ij}^{a0,a\mu }}}{{{\Delta _{a\mu }}}}} \right\} \sim \left( {w_a^ + w_a^ - } \right)\sum\limits_\mu  {\frac{{{{\left( {t_{ij}^{a0,b\mu }} \right)}^2}}}{{{\Delta _{{b}a\mu }}}}} }
%\label{eq:26}
%\end{eqnarray}
%\end{widetext}
%\begin{widetext}
%\begin{eqnarray}
%J_{ij}\sim0.146eV+\left(w^+_a+w^-_a\right)\times...+\left( {w_a^+ w_a^- } \right)\times...
%\left\langle {\delta {H_{s - ex}}} \right\rangle  =...
%\label{eq:27}
%\end{eqnarray}
%\end{widetext}
What are the modifications of the exchange interaction that we can observe in L214 under resonance light pumping? The answer to this question depends on the ratio of the exchange interaction in the ground and excited states. Depletion of the ground state $p_b=1-x$ decreases $J_{bb}$, and a new contribution $J_{ba}$ via excited orbital $a_1$ appears (see Fig.\ref{fig:2}). Using LDA parameters, and summing over all $\mu$ in the second term in Eq.(\ref{eq:25}), we finally obtain the result shown in Fig.\ref{fig:3}. So most likely superexchange contribution (\ref{eq:25}) will increase at any small population of excited states in La214 by a factor of $ \sim~ 4\cdot10^{-3} eV(\%)^{-1}$

\section{\label{sec:IV} Conclusion}

In summary, we would like to emphasize that optical pumping results in the occupation of some high energy multielectron states with different overlapping of the excited wavefunctions between neighboring ions vs the ground state orbitals. It is evident that this pumping results in the modification of the exchange interaction. Nevertheless an accurate calculation of a large number of contributions from different multielectron excited states is not a trivial theoretical problem. The gain of the Hubbard operators approach is the ability to control each excited state and its contribution to the ionic spin and orbital moment. Our approach to the exchange interaction via excited states is just a straightforward generalization of the previously  developed projection technique for the Hubbard model~\cite{Chao_etal1977} and for the ground state of La214 within the realistic multiband $pd$ model.~\cite{Gavrichkov_etal2008}
The  obtained effective Hamiltonian (\ref{eq:19}) contains not only spin-spin interactions via excited states but also more complicated exchange interactions accompanied with exciton or bi-exciton that are beyond standard  Heisenberg model.

For  undoped insulating cuprates the theory results in a prediction of the antiferromagnetic coupling strengthening proportional to the concentration of the excited states
At the concentration of excited states 1\% an increased exchange interaction is estimated by the magnitude $\sim 40$K.
For simplicity we have assumed stationary pumping with resonance absorbtion. Then the spectral dependence of the modified exchange coupling should coincide with the $d-d$ absorption spectrum. Due to the short time of the local electronic excitations  $\leq$ 1 (fs) a dynamics of exchange interaction for the time intervals more then 10 (fs) probably can be also treated in our approach. It is evidently that the spin-exciton effects found here may be important in the dynamical regimes.

\begin{acknowledgments}
This work was supported by RFFI grants 16-02-00273, No.14-02-00186.
\end{acknowledgments}

\bibliography{my}
\end{document}